\documentclass[prd,floatfix,preprintnumbers,letterpaper]{revtex4}
\usepackage{graphicx}
\usepackage{amsmath,amssymb}
\input{epsf}
\usepackage{epsf}
\usepackage{subfigure}
\usepackage{color}
\usepackage{hyperref}
\usepackage{float}

\newcommand {\ga} {\ {\raise-.5ex\hbox{$\buildrel>\over\sim$}}\ }
\newcommand {\la} {\ {\raise-.5ex\hbox{$\buildrel<\over\sim$}}\ } 

\begin{document}

\title{Reconstructing the slope of a nearly flat quintessence potential from cosmography}

\author{Saikat Chakraborty}
\affiliation{Institute of Research and Development, Duy Tan University, Da Nang 550000, Vietnam}
\affiliation{Faculty of Natural Sciences, Duy Tan University, Da Nang 550000, Vietnam}
\affiliation{Centre for Space Research, North-West University, Potchefstroom 2520, South Africa}
\email{saikat.chakraborty@nwu.ac.za, peter.dunsby@uct.ac.za, robert.scherrer@vanderbilt.edu}

\author{Peter K.S. Dunsby}
\affiliation{Department of Mathematics and Applied Mathematics, University of Cape Town, Rondebosch 7701, Cape Town, South Africa}
\affiliation{South African Astronomical Observatory, Observatory 7925, Cape Town, South Africa}
\affiliation{Centre for Space Research, North-West University, Potchefstroom 2520, South Africa}

\author{Robert J. Scherrer}
\affiliation{Department of Physics and Astronomy, Vanderbilt University,
Nashville, TN 37235}

\begin{abstract}
We revisit thawing quintessence models with nearly flat scalar-field potentials using a cosmographic framework. 
Earlier work indicates that the cosmographic reconstruction of the slope $\lambda=-(dV/d\phi)/V$ of the quintessence potential in the general case requires the knowledge of the cosmographic paremeters up to the jerk parameter $j$.  In this work we show that the slow-roll conditions $[(dV/d\phi)/V]^2 \ll 1$ and $|(d^2V/d\phi^2)/V| \ll 1$ allow the reconstruction of the slope of a nearly flat potential with knowledge of only the deceleration parameter $q$ (and the density parameter $\Omega_\phi$). Confronting the assumption of near-flatness with the cosmographic data after DESI DR2, however, reveals possible tension between the two.
We further show that these models exhibit attractor behaviour in the $w$--$\Omega_\phi$ and $w$--$w'$ phase planes, corresponding to a universal thawing evolution with $w \approx -1$ at early times. We also derive the corresponding relation in the cosmographic $q$--$j$ plane and show that different cosmological expansion histories can produce the same thawing evolution. Nevertheless, all viable trajectories remain close to the $\Lambda$CDM limit $j=1$.
\end{abstract}

\maketitle
\section{Introduction}

The origin of the accelerated expansion of the Universe \cite{Riess,Perlmutter} remains one of the central open problems in modern cosmology. Within the standard cosmological paradigm, this acceleration is described by the $\Lambda$CDM model, in which a cosmological constant acts as a constant vacuum energy density that drives late-time cosmic acceleration. Nevertheless, present observations do not exclude the possibility that dark energy is dynamical in nature. In particular, recent measurements from the Dark Energy Spectroscopic Instrument (DESI) have suggested possible departures from a pure cosmological constant \cite{DESIDR1,DESIDR2,DESIDR2p2}, thereby renewing interest in models where the dark energy density evolves with cosmic time.

One of the most extensively studied classes of dynamical dark energy models is quintessence, in which the accelerated expansion is driven by a scalar field $\phi$ evolving in a potential $V(\phi)$ \cite{RatraPeebles,Wetterich,Ferreira,CLW,Ferreira2,CaldwellDaveSteinhardt,Liddle,SteinhardtWangZlatev}; see Ref.~\cite{Copeland1} for a review. These models are commonly characterized by the equation-of-state parameter $w \equiv p/\rho$, defined as the ratio of the dark energy pressure to its energy density. A cosmological constant corresponds to $w=-1$, yielding a constant energy density, whereas models with $w>-1$ generally predict an evolving dark energy component.

Although the parameter $w$ provides a convenient phenomenological description, it is not directly observable. An alternative and largely model-independent approach is provided by cosmography, in which the expansion history of the Universe is characterized through successive derivatives of the scale factor $a$ \cite{Sahni}. In this framework, cosmic expansion is described by kinematic quantities such as the Hubble parameter $H \equiv \dot{a}/a$, the deceleration parameter $q \equiv -\ddot{a}/(aH^2)$, and higher-order parameters including the jerk $j \equiv \dddot{a}/(aH^3)$ and the snap $s \equiv \ddddot{a}/(aH^4)$. While this hierarchy can, in principle, be extended indefinitely, only the lowest-order cosmographic parameters are presently accessible observationally (see Ref.~\cite{Dunsby} for a review).

A number of studies have investigated the relationship between cosmographic (or statefinder) parameters and quintessence dynamics \cite{Nair,Dubey,Gerke,Sahlen,Mukherjee,Elizalde}. In Ref.~\cite{paper1}, Chakraborty et al.\ derived exact expressions for the first and second derivatives of the quintessence potential in terms of the scalar-field density parameter $\Omega_\phi$ and the cosmographic parameters $q$, $j$, and $s$. Although these expressions are independent of an explicit parameterization in terms of $w$, they are algebraically cumbersome and involve higher-order cosmographic quantities that remain only weakly constrained by current observations.

In the present work, we extend these results to a particularly well-motivated subclass of quintessence models: thawing quintessence with a nearly flat potential. Such models, studied in detail in Ref.~\cite{ScherrerSen}, satisfy the conditions $[(dV/d\phi)/V]^2 \ll 1$ and $|(d^2V/d\phi^2)/V| \ll 1$. In these scenarios, the scalar field is initially frozen by Hubble friction, so that $w \approx -1$, and the scalar field only begins to evolve at late times as it gradually ``thaws'' and rolls down the potential. Consequently, the equation-of-state parameter departs slowly from $-1$ while remaining close to it throughout the evolution. An important feature of these models is the emergence of a universal evolutionary behaviour, encapsulated in a simple relation between $w$, $\Omega_\phi$, and the initial value of $(dV/d\phi)/V$. Here, we extend the formalism developed in Ref.~\cite{paper1} to this class of thawing quintessence models. We adopt units where $8 \pi G= c=1$.

\section{Quintessence model}
In this section, we briefly review the quintessence framework and introduce the dynamical system that underpins our subsequent analysis. We consider dark energy to be described by a minimally coupled scalar field, $\phi$, evolving in a potential $V(\phi)$. The pressure and energy density associated with the scalar field take their standard forms,
\begin{eqnarray}
p &=& \frac{\dot \phi^2}{2} - V(\phi),\\
\rho &=& \frac{\dot \phi^2}{2} + V(\phi),
\end{eqnarray}
where an overdot denotes differentiation with respect to cosmic time. Conservation of energy-momentum then yields the scalar-field equation of motion,
\begin{equation}
\label{motionq}
\ddot{\phi}+ 3H\dot{\phi} + \frac{dV}{d\phi} = 0,
\end{equation}
with the Hubble expansion rate given by
\begin{equation}
\label{H}
H = \sqrt{\rho_{\rm tot}/3},
\end{equation}
where $\rho_{\rm tot}$ denotes the total cosmic energy density, consisting of non-relativistic matter and the scalar-field contribution.


The cosmological field equations of the quintessence model can be conveniently expressed as the following autonomous system \cite{ScherrerSen}
\begin{subequations}\label{autonomous}
\begin{align}
\frac{dw}{d\ln a} &= \lambda(1-w)\sqrt{3(1+w)\Omega_\phi}\,-3\left(1-w^2\right)\,,\label{w'-eq}
\\
\frac{d\Omega_{\phi}}{d\ln a} &= -3w\Omega_\phi(1-\Omega_\phi)\,,\label{Omega'-eq}
\\
\frac{d\lambda}{d\ln a} &= -\sqrt{3}(\Gamma-1)\lambda^2\,\sqrt{(1+w)\Omega_\phi}\,,\label{lambda'-eq}
\end{align}
\end{subequations}
where the derivatives of the potential
are expressed in terms of
\begin{equation}
\lambda \equiv - \frac{1}{V}\frac{dV}{d\phi},
\end{equation}
and
\begin{equation}
\Gamma \equiv V \frac{d^2V}{d\phi^2}/\left(
\frac{dV}{d\phi}\right)^2.
\end{equation}

It is worth pointing out that the autonomous system corresponding to the quintessence model can also be written in an interesting new way: 
\begin{subequations}\label{autonomous_new}
\begin{align}
   \frac{dw}{d\ln a}=& ~w'\,,
   \\
   \frac{dw'}{d\ln a} =& ~6\,w\,w' + (3-3w^2+w')\bigg[\frac{1-\Gamma-w'}
{1-w}\nonumber\\
   & +\frac{w\,w'^2+3w(1+w)(1-w)^2(2+3w-\lambda)^2+(1-w)w'\lbrace6w(1+w)+(1-w)\lambda^2\rbrace}{2(1-w)^2(1+w)\lambda^2}\bigg]\,,
   \\
   \frac{d\lambda}{d\ln a} =& ~\frac{(3-3w^2+w')(1-\Gamma)\lambda}{1-w}\,.
\end{align}
\end{subequations}
where prime denotes derivative with respect to $\ln a$.
The above form of the autonomous system is particularly useful to visualize the dynamics in the phenomenological $w-w'$ plane as we demonstrate below.

Our goal is to express the potential in terms of
the cosmographic parameters, the first of which are
the deceleration parameter $q$ and the jerk parameter $j$. These are related to the
physical parameters $w,\,w'$ and $\Omega_{\phi}$ as follows
\begin{subequations}
\begin{eqnarray}
\label{q}
q &=& \frac{1}{2} + \frac{3}{2} w \Omega_\phi,\\
\label{j}
j &=& 1 + \frac{3}{2} \Omega_\phi(3w +3w^2 - w^\prime).
\end{eqnarray}
\end{subequations}
In Ref. \cite{paper1} it was shown that $\lambda$
can be expressed as a function of $\Omega_\phi$, $q$, and $j$.  Here we seek a simpler expression for the special case of thawing quintessence in a nearly flat potential.

\section{Thawing Quintessence with Nearly Flat Potential}\label{sec:recon_lambda}

Consider the model discussed in Ref. \cite{ScherrerSen}, for a scalar field evolving in a nearly flat potential. Following \cite{ScherrerSen}, the conditions for sufficient flatness around an initial value $\phi=\phi_0$ are taken to be
\begin{equation}
\label{slow1}
\left(\frac{1}{V} \frac{dV}{d\phi}\right)^2 \bigg\vert_{\phi=\phi_0} \ll 1,
\end{equation}
and
\begin{equation}
\label{slow2}
\frac{1}{V}\frac{d^2 V}{d\phi^2} \bigg\vert_{\phi=\phi_0} \ll 1.
\end{equation}
When these conditions are satisfied, and the field is
initially frozen with $w =-1$, it can be shown that
$w\approx -1$ at all later times, and $\lambda$ is roughly constant with a value of $\lambda_0$.
In this case, Eqs. \eqref{w'-eq} and \eqref{Omega'-eq} can be combined to give \cite{ScherrerSen}
\begin{equation}
    \frac{dw}{d\Omega_{\phi}} = -\frac{2(1+w)}{\Omega_{\phi}(1-\Omega_{\phi})} + \frac{2}{3}\lambda_0\frac{\sqrt{3(1+w)}}{(1-\Omega_{\phi})\sqrt{\Omega_{\phi}}}\,.
\end{equation}
With the initial boundary condition $w=-1$ at $\Omega_\phi = 0$, this equation admits the solution
\cite{ScherrerSen}
\begin{eqnarray}
1 + w &=& \frac{\lambda_0^2}{3}\left[\frac{1}{\sqrt{\Omega_\phi}}
- \left(\frac{1}{\Omega_\phi} - 1 \right) \tanh^{-1}\sqrt{\Omega_\phi}\right]^2,\nonumber\\
\label{attractor_1}
&=& \frac{\lambda_0^2}{3}\left[\frac{1}{\sqrt{\Omega_\phi}}
- \frac{1}{2}\left(\frac{1}{\Omega_\phi} - 1 \right)
\ln \left(\frac{1+\sqrt{\Omega_\phi}}
{1-\sqrt{\Omega_\phi}} \right)\right]^2,
\end{eqnarray}
where the assumption $\lambda_0^2 \ll 1$ ensures that $w \approx -1$. It was shown in Ref. \cite{ScherrerSen} that this solution acts as an attractor solution, i.e., a common evolutionary track for nearly flat potentials.

We can use Eq. \eqref{w'-eq} to eliminate
$\Omega_\phi$ from  \eqref{attractor_1} in favor
of $w^\prime$, giving an alternative expression for the attractor solution,
in the $w-w'$ plane:
\begin{equation}\label{attractor_2}
1+w = \frac{\lambda_0^2}{3}\left[\frac{\sqrt{3}\lambda_0(1-w)\sqrt{1+w}}{3-3w^2+w'} - \left(\frac{3\lambda_0^2 (1-w)^2 (1+w)}{(3-3w^2+w')^2}-1\right) \tanh^{-1}\left(\frac{(3-3w^2+w')}{\sqrt{3}\lambda_0(1-w)\sqrt{1+w}}\right)\right]^2 
\end{equation}

\begin{figure}[H]
\centering
\includegraphics[width=\linewidth]{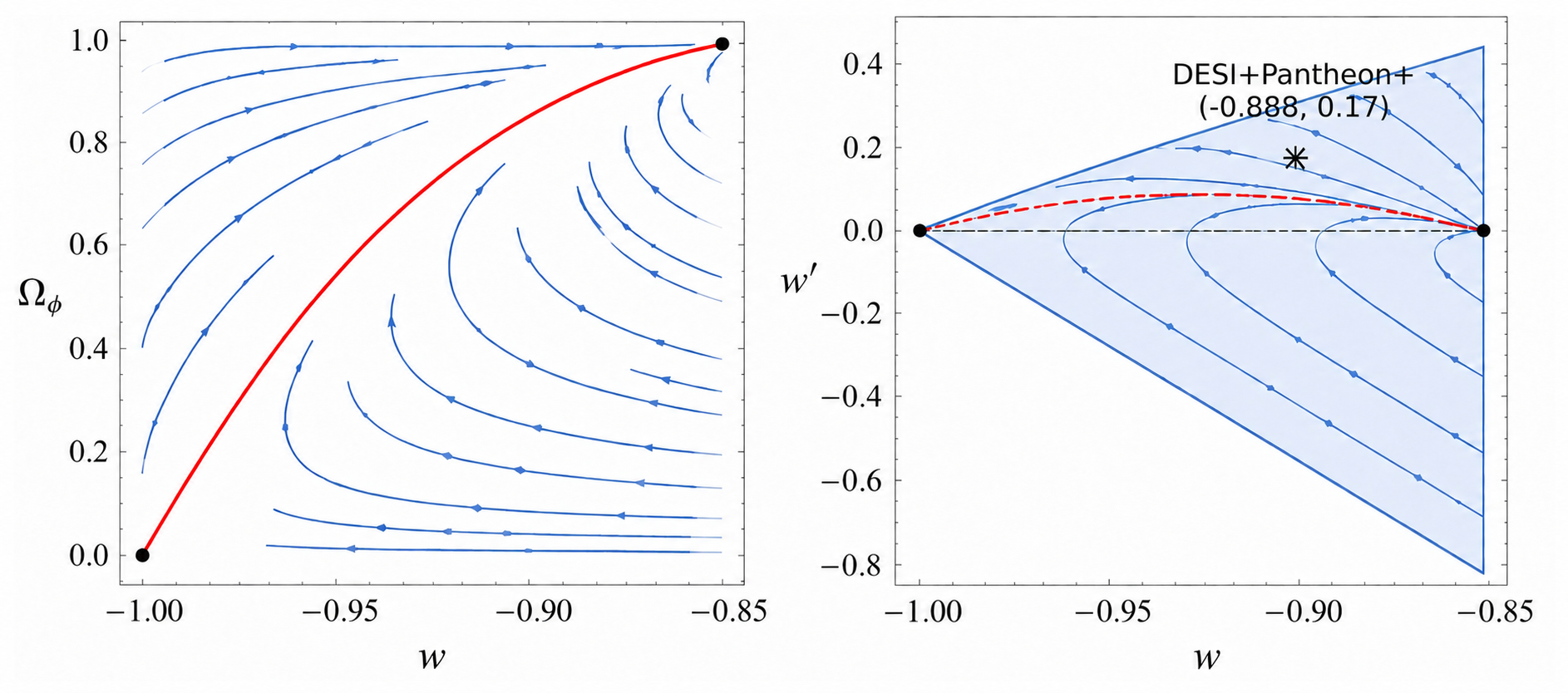}
\caption{The 2D phase portraits corresponding to a nearly flat potential for $\lambda\approx\lambda_0=\frac{2}{3}$ (which has been chosen for illustrative purposes, leading to a clear visual separation between trajectories), with fixed points shown as black dots. The attractor solutions \eqref{attractor_1} and \eqref{attractor_2} are shown in the respective figures as red curves. The attractor solution connects the matter-dominated fixed point $(w,\Omega_{\phi})=(-1,0)$ (or $(w,w')=(-1,0)$) to the quintessence dominated fixed point $(w,w')=\left(-1+\lambda_0^2/3,1\right)$ (or $(w,w')=(-1+\lambda_0^2/3,0)$). In the right panel, the shaded region corresponds to $-3(1-w^2)<w'<3(1+w)$, which is a bound on $w'$ for thawing quintessence; see Ref. \cite{Scherrerww'}. Also shown in the right panel is the DESI + Pantheon+ value of $(w,w')_{z=0}$ taken from \cite[Table V]{DESI:2025zgx}, which appears quite close to the attractor solution.}
    \label{fig:attractor}
\end{figure}

Under the assumption of a roughly constant $\lambda$, with $\lambda\approx\lambda_0$ and $\Gamma\approx1$, the effective 2-dimensional phase portraits of the phase space in the $w-\Omega_{\phi}$ plane and $w-w'$ plane are shown in Fig. \ref{fig:attractor}. 
The attractor solution is also explicitly shown in the figure. Different phase trajectories in the phase portrait correspond to different possible evolutionary paths of the cosmological system, or different possibilities for the dark energy evolution. Convergence of the trajectories towards the attractor solution signifies convergence towards a common evolutionary tract of the dark energy equation of state. The attractor solution in the $w-w'$ plane, in particular, shows that $w$, initially frozen at the value $-1$, thaws with the cosmic evolution and stabilizes at the value $w=-1+\frac{1}{3}\lambda_0^2$, which corresponds to Eq. (\ref{attractor_1}) in the limit $\Omega_\phi \rightarrow 1$.

We now use Eq. (\ref{q}) to eliminate $w$ from Eq.
(\ref{attractor_1}), giving
\begin{equation}
\label{lambda_0}
\lambda_0 = \frac{\sqrt{3 \Omega_{\phi} + 2q -1}}{1 - \left(\frac{1}{\sqrt{\Omega_{\phi}}}- \sqrt{\Omega_{\phi}}\right)
\tanh^{-1}\sqrt{\Omega_\phi}}
\end{equation}
Eq. (\ref{lambda_0}) is the main result of this paper. Note that our derivation implies that it is valid at all times up to the present, despite the fact that both $q$ and $\Omega_{\phi}$ evolve with time.  Practically speaking, $q$ can only be measured accurately at the present, so a determination of $\lambda_0$ requires us to take $q$ and $\Omega_\phi$ equal to their present-day values.

Unlike the case of exact evolution studied in Ref. \cite{paper1}, our expression for $\lambda_0$ is independent of $j$.
This is a major advantage, as $j$ is much more poorly measured than $q$. Indeed this result suggests a broader cosmographic reconstruction programme for dynamical dark energy. While the exact relations derived in Ref. \cite{paper1} connect the first and second derivatives of the quintessence potential to the cosmographic parameters $(q,j,s)$, the simplified relation obtained here demonstrates that, in the slow-roll thawing limit, useful information about the underlying potential may already be extracted from lower-order cosmographic observables. Future work may therefore seek to combine these exact and approximate reconstruction schemes in order to recover both the slope and curvature of the quintessence potential directly from observations.

Table \ref{tab:cosm_values_dr2} lists the best-fit present-day value of the deceleration parameter $q_0$, found using Taylor series cosmography, Pad\'e cosmography, and Chebyshev cosmography applied to DESI DR2 data (see \cite[Table 1]{cosmography_DESI2}). The respective values of the potential slope parameter $\lambda_0$, as calculated from Eq.\eqref{lambda_0} with a fiducial value $\Omega_{\phi0}=0.7$, are also listed.
\begin{table}[H]
    \centering
    \begin{tabular}{|c|c|c|}
    \hline 
        Method & $q_0$ & $\lambda_0$
        \\
        \hline 
        Taylor ($z\leq1$) & $-0.41$ & $0.935$ 
        \\
        \hline 
        Pad\'e$_{(2,1)}$ & $-0.49$ & $0.612$
        \\
        \hline 
        Pad\'e$_{(2,2)}$ & $-0.397$ & $0.977$
        \\
        \hline 
        Chebyshev & $-0.472$ & $0.698$ 
        \\
        \hline
    \end{tabular}
    \caption{Cosmographic values taken from \cite[Table 1]{cosmography_DESI2}. $\Omega_{\phi0}=0.7$ is assumed.}
    \label{tab:cosm_values_dr2}
\end{table}
 The values shown in Table \ref{tab:cosm_values_dr2} suggest that the inferred slope parameter may be moderately sensitive to the adopted cosmographic reconstruction method. This sensitivity motivates a more comprehensive likelihood analysis using DESI, Pantheon+, and related datasets, with the aim of deriving robust confidence intervals on $\lambda_0$ and determining whether present observations favour potentials that satisfy the slow-roll conditions assumed in this work. Irrespective of the adopted cosmographic method, however, some of the values in Table \ref{tab:cosm_values_dr2} suggest that the DESI DR2 data might be in tension with the assumption of near flatness. From Eq. (\ref{attractor_1} we can derive a somewhat more precise condition on $\lambda_0$ for the slow-roll condition to hold, namely $\lambda_0^2/3 \ll 1$. The two smaller values of $\lambda_0$ in Table \ref{tab:cosm_values_dr2} satisfy this constraint, while the other two do not.
 
\begin{figure}[H]
\centering
\includegraphics[width=0.5\linewidth]{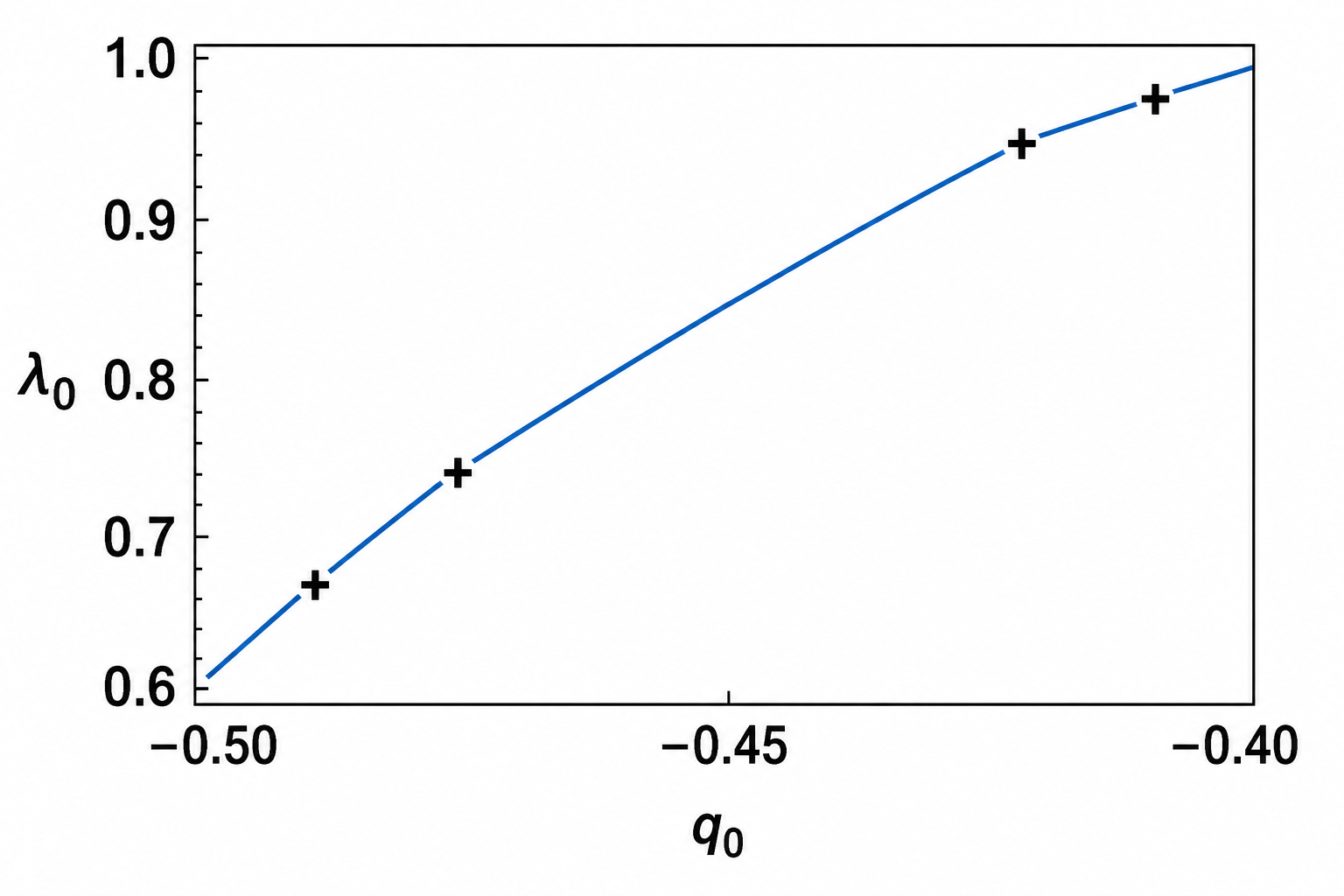}
\caption{The plot of $\lambda_0$ vs $q_0$ as obtained from the relation \eqref{lambda_0}, with $\Omega_{\phi0}=0.7$. The values obtained in Table \ref{tab:cosm_values_dr2} are marked with a $``+''$ sign. }
    \label{fig:lambda0_vs_q0}
\end{figure}

Fig.\ref{fig:lambda0_vs_q0} shows a plot of $\lambda_0$ vs $q_0$ as obtained from the relation \eqref{lambda_0}, with $\Omega_{\phi0}=0.7$. As is apparent from the graph, the derived value of $\lambda_0$ is quite sensitive to small changes in the estimated value of $q_0$.
For a variation of $q_0$ from $-0.5$ to $-0.4$, $\lambda_0$ varies from $0.6$ to $1.0$, and for $q_0$ in this range, the value of $d\lambda_0/dq_0$ ranges from $5.6$ to $3.2$, decreasing as $q_0$ increases toward 0.  An accelerating universe ($q_0 < 0$) with $\Omega_{\phi 0} = 0.7$ requires
a value of $\lambda_0 < 1.9$.
The strong dependence of $\lambda_0$ on $q_0$ suggests that future improvements in cosmographic measurements may lead to substantially tighter constraints on the slope of the quintessence potential. In this sense, Eq. (15) provides a direct observational target for forthcoming surveys such as Euclid, Roman, Rubin Observatory and the SKA.

\section{Dynamics under nearly flat potential in the cosmographic plane $q-j$}

In this section, we examine the cosmological dynamics under a thawing quintessence model with a nearly flat potential in the cosmographic plane $q-j$ (which is also called the statefinder plane \cite{Sahni,Nair,Dubey}). First note that the dynamical equation for $\Omega_{\phi}$, namely Eq.\eqref{Omega'-eq}, can be expressed as
\begin{equation}\label{Omega'-eq_new}
    \frac{d\Omega_{\phi}}{d\ln a} = (1-2q)(1-\Omega_{\phi})\,,
\end{equation}
where we have used the relation \eqref{q}. The deceleration and the jerk parameter themselves are related as
\begin{equation}\label{q'-eq} 
  j = 2q^{2} + q - \frac{dq}{d\ln a}\,.
\end{equation}
Now, taking the derivative of the expression \eqref{lambda_0}, and using the relations \eqref{Omega'-eq_new} and \eqref{q'-eq}, we get the following relation
\begin{equation}\label{attractor_3}
    j-1 = -(1+q)(1-2q) + (1-2q)(1-\Omega_{\phi})\left[\frac{3}{2} + \frac{\lambda_0^2}{2\Omega_{\phi}} - \frac{\lambda_0^2}{2\Omega_{\phi}^{3/2}}\ln\left(\frac{1+\sqrt{\Omega_{\phi}}}{1-\sqrt{\Omega_{\phi}}}\right) + \frac{\lambda_0^2}{8}\left(\frac{1}{\Omega_\phi^2}-1\right)\ln\left(\frac{1+\sqrt{\Omega_{\phi}}}{1-\sqrt{\Omega_{\phi}}}\right)^2\right]\,.
\end{equation}

A notable feature of this result is that the attractor solution (Eqs. \ref{attractor_1} or \ref{attractor_2}) does not correspond to a single trajectory in the $q-j$ plane. Instead, it maps to a one-parameter  ($\Omega_{\phi}$) family of curves. Since a particular curve in the $q-j$ plane implies a particular kind of evolutionary history $a(t)$, this result indicates that different kinds of evolutionary histories can still produce the same kind of evolution of the dark energy equation of state parameter. This is an affirmation of the fact that the cosmographic observables such as $q$ and $j$ do not uniquely determine the underlying dark energy dynamics. This degeneracy should be taken into account when interpreting observational constraints on the evolution of the dark energy equation of state and highlights the fact that a reconstruction of the dark-energy equation of state does not uniquely determine the underlying expansion history.

However, all the possible evolutionary histories given by \eqref{attractor_3}, even though different, are still close to $j=1$, which is the cosmographic specification of the $\Lambda$CDM model in which $w=-1$ \cite{Sahni}. To establish that, we substitute $\Omega_{\phi}=\frac{2q-1}{3w}$ on the right-hand side of \eqref{attractor_3}, and expand around $w=-1$ to the lowest order. We obtain
\begin{subequations}
\begin{align}
& j-1 = -\frac{1}{2}(1+w)(2q-1)^2 + \lambda_0^2 (1+q) \left[2q-1 + \frac{(q-2)(q+1)\log^2\left(\frac{3-\sqrt{3-6q}}{3+\sqrt{3-6q}}\right) + 3\sqrt{3-6q}\log\left(\frac{3-\sqrt{3-6q}}{3+\sqrt{3-6q}}\right)}{3(2q-1)}\right]\nonumber\\
& -(1+w)\lambda_0^2\left[3 - \log\left(\frac{3-\sqrt{3-6q}}{3+\sqrt{3-6q}}\right)\frac{(2q^2-5q-16)\sqrt{3-6q}+(1+q)(2q^2-5q+11) \log \left(\frac{3-\sqrt{3-6q}}{3+\sqrt{3-6q}}\right)}{6(2q-1)}\right]\nonumber\\
& + \mathcal{O}[(1+w)^2]\,.
\end{align}
\end{subequations}
Since a nearly flat potential requires satisfying the slow-roll condition $\lambda_0^2 \ll 1$ (Eq.\eqref{slow1}), $w\approx-1$ implies $j\approx1$.

The above result is an expected one. Eqs.\eqref{q} and \eqref{j} can be combined to give
\begin{equation}
    j-1 = 3\left(q-\frac{1}{2}\right)\left(1+w-\frac{w'}{3w}\right)\,.
\end{equation}
Together, the conditions \eqref{slow1} and \eqref{slow2} for a sufficiently flat potential imply that $w$ remains close to $-1$ for the entire course of evolution. From the above expression, $(w,w')\approx(-1,0)$, which implies $j\approx1$.

\section{Conclusions}

In this work, we have investigated thawing quintessence models with nearly flat scalar-field potentials within a cosmographic framework. By imposing the slow-roll conditions 
\[
\left( \frac{1}{V}\frac{dV}{d\phi} \right)^2 \ll 1
\quad \text{and} \quad
\left| \frac{1}{V} \frac{d^2 V}{d\phi^2}\right| \ll 1 ,
\]
we derived a simplified expression for the potential slope parameter $\lambda_0$ in terms of the dark-energy density parameter $\Omega_\phi$ and the deceleration parameter $q$. Unlike the exact expressions previously obtained for generic quintessence evolution, the resulting relation is independent of the jerk parameter $j$, which remains poorly constrained observationally. This provides a considerably more practical connection between cosmographic observables and scalar-field dynamics.

We further showed that thawing quintessence with nearly flat potentials exhibits attractor behaviour in both the $w$--$\Omega_\phi$ and $w$--$w'$ phase planes. Independent of the detailed form of the potential, the scalar field evolves along a common trajectory characterized by $w \approx -1$ at early times, followed by a gradual thawing at late epochs. The corresponding phase portraits illustrate the universality of this evolution and clarify the dynamical interpretation of the attractor solutions derived in Ref.~\cite{ScherrerSen}.

We also examined the dynamics in the cosmographic $q$--$j$ plane. In contrast to the $w$--$\Omega_\phi$ and $w$--$w'$ descriptions, the attractor solution does not map to a unique trajectory in the statefinder plane, but rather to a family of curves parameterized by $\Omega_\phi$. This demonstrates that distinct cosmological expansion histories may lead to the same effective thawing behaviour of the dark-energy equation-of-state parameter. Nevertheless, all viable trajectories remain close to the $\Lambda$CDM limit $j = 1$, reflecting the fact that nearly flat thawing models remain observationally close to a cosmological constant.

The relations derived here are formally valid at arbitrary redshift, provided the evolution of $\Omega_\phi$ is consistently taken into account. In practice, however, the framework is most useful near the present epoch, where cosmographic parameters can be constrained observationally with reasonable precision. Future observational improvements, particularly in measurements of the late-time expansion history, may allow these relations to serve as a direct probe of the slope of the quintessence potential.

Several extensions of the present work would be worthwhile. In particular, it would be interesting to confront Eq. \eqref{lambda_0} with current DESI and supernova observations in order to derive direct constraints on the slope of the quintessence potential. More generally, the present analysis suggests that cosmographic observables may provide a useful bridge between observable expansion history and the underlying scalar-field dynamics, opening the possibility of increasingly detailed reconstructions of dark-energy potentials from future surveys.

\acknowledgments
PKSD thanks First Rand Bank (SA) for financial support.

\end{document}